\documentclass[preprint,aps,showpacs,amsmath]{revtex4}

\usepackage{dcolumn}
\usepackage{bm}

\newcommand*{\de}{\,{\rm d}}
\renewcommand*{\vec}[1]{{\rm\bf #1}}
\newcommand*{\frct}[2]{{\textstyle\frac{#1}{#2}}}
\newcommand*{\ka}{\kappa}
\newcommand*{\la}{\lambda}
\newcommand*{\vp}{\vphantom{_0^0}}

\begin{document}

\title{Combination of many-body and density-functional theories}

\author{Dimitri N. Laikov}
\email{laikov@physto.se}
\affiliation{Department of Physics, AlbaNova University Center,
Stockholm University, SE-10691 Stockholm, Sweden}

\date{\today}

\begin{abstract}
A framework for developing new approximate electronic structure methods
is presented, in which the correlation energy of a many-electron system
in the ground state is computed as in the single-reference
 second-order many-body perturbation
theory but with the reference one-body Hamiltonian modified by additional
(generally non-local) potentials which are universal functionals
of the Hartree-Fock density.
The existence of such functionals that reproduce the exact
correlation energy justifies the search for approximate models
which may overcome some important deficiencies of the traditional
density-functional methods and also perform better than the usual
low-order wavefunction methods.
\end{abstract}

\pacs{31.10.+z, 31.15.Ew, 31.15.Md}

\maketitle

\section{Introduction}

The Hohenberg-Kohn theorems~\cite{HK64} have provided a formal justification
for the development of a number of approximate density functionals
for electronic structure calculations within the Kohn-Sham scheme~\cite{KS65}.
Such density-functional methods are now widely used in theoretical chemistry
for computing ground state properties of molecules
due to their low cost and scalability.
Nevertheless, there is still a number of deficiencies that the traditional
approximate functionals cannot overcome.
Self-interaction error~\cite{PZ81} is one such
example, another important problem is the inability to describe the
dispersion (van der Waals) interaction~\cite{KMM98}.
On the other hand, the simplest correlated wavefunction
method -- the M{\o}ller-Plesset second-order (MP2) many-body
perturbation theory~\cite{MP34} --
does not suffer from these deficiencies, its cost, though higher, is still
acceptable for many interesting applications. There are cases, however,
where this simple approximation shows its weakness --
for strongly-correlated systems the correlation energy can be significantly
overestimated, another difficulty appears when the underlying Hartree-Fock
reference experiences an artificial spin- or simmetry-breaking.

In this work we propose a framework for developing new approximated methods
of electronic structure calculations which combine the strengths
of both the low-order many-body perturbation theory and
the local density functional approximations.

\section{Theory}

Starting from the fact that the exact total energy of a many-electron system
in the ground state
is a functional of the Hartree-Fock density~\cite{D90}, we propose here
a special form of the correlation energy functional for which simple
approximate models may be developed.

The single-determinat wavefunction $\Phi$ of the Hartree-Fock theory
assumed normalized
\begin{equation}
\left<\Phi\right|\left.\!\Phi\right> = 1
\end{equation}
makes the expectation value of the energy
\begin{equation}
\label{E0}
E_0 = \left<\Phi\right|\hat{H}\left|\Phi\right>
\end{equation}
stationary, which is equivalent to the conditions
\begin{equation}
\label{Fia}
\left<\Phi_i^a\right|\hat{H}\left|\Phi\right> = 0,
\end{equation}
where singly-substituted determinants $\Phi_i^a$ are introduced.
The indices $i,j,k,\dots$ ($a,b,c,\dots$) label the occupied (virtual)
one-electron wavefunctions, whereas $\ka,\la,\mu,\nu$ will refer to any of them.

In terms of one- and two-electron integrals
\begin{equation}
H_\mu^\nu =
\int \phi_\mu(\vec{r})\left(-\frct12\nabla^2 + v(\vec{r})\right)\phi_\nu(\vec{r})
 \de^3\vec{r},
\end{equation}
\begin{equation}
R_{\ka\mu}^{\la\nu} =
\bar{R}_{\ka\mu}^{\la\nu} - \bar{R}_{\ka\mu}^{\nu\la},
\end{equation}
\begin{equation}
\bar{R}_{\ka\mu}^{\la\nu} =
\int \phi_\ka(\vec{r})\phi_\mu(\vec{r'}) |\vec{r}-\vec{r'}|^{-1}
     \phi_\la(\vec{r})\phi_\nu(\vec{r'}) \de^3\vec{r} \de^3\vec{r'},
\end{equation}
where the wavefunctions $\phi_\ka(\vec{r})$ are assumed to be real and their
spin dependence is not show explicitly for simplicity,
the energy~(\ref{E0}) and the conditions~(\ref{Fia}) take the form
\begin{equation}
E_0 = H_i^i + \frct12 R_{ij}^{ij},
\end{equation}
\begin{equation}
F_i^a = 0,
\end{equation}
with the Fock matrix defined by
\begin{equation}
F_\mu^\nu = H_\mu^\nu + R_{\mu i}^{\nu i}.
\end{equation}
The summation over all repeated indices is assumed throughout.

We now introduce a first-order correlated wavefunction $\Psi_1$ as
a linear combination of all doubly substituted determinants
\begin{equation}
\Psi_1 = \frct14 \tau_{ij}^{ab} \Phi_{ij}^{ab},
\end{equation}
from which the (exact) correlation energy will be computed as
\begin{equation}
\label{Ec}
E_{\rm c} = \left<\Psi_1\right|\hat{H}\left|\Phi\right>
\end{equation}
or in terms of two-electron integrals
\begin{equation}
\label{Eijab}
E_{\rm c} = \frct14 \tau_{ij}^{ab} R_{ij}^{ab}.
\end{equation}
The coefficients $\tau_{ij}^{ab}$ are determined by linear equations
\begin{equation}
\label{tijab}
  f_c^a \tau_{ij}^{cb} + f_c^b \tau_{ij}^{ac}
- f_i^k \tau_{kj}^{ab} - f_j^k \tau_{ik}^{ab}
+ R_{ij}^{ab} = 0
\end{equation}
which are formally derived from many-body perturbation theory
with the one-electron matrix elements of the zeroth-order Hamiltonian
in the form
\begin{eqnarray}
f_i^j &=& F_i^j + u_i^j, \\
f_a^b &=& F_a^b + u_a^b,
\end{eqnarray}
\begin{eqnarray}
u_i^j &=&
 \int\phi_i(\vec{r})u_{\rm o}(\vec{r};[\rho_0])\phi_j(\vec{r}) \de^3\vec{r}, \\
u_a^b &=&
 \int\phi_a(\vec{r})u_{\rm v}(\vec{r};[\rho_0])\phi_b(\vec{r}) \de^3\vec{r}.
\end{eqnarray}
The two different ``correction potentials" $u_{\rm o}$ and $u_{\rm v}$
to be added to the one-electron Fock operator are universal functionals
of the Hartree-Fock density
\begin{equation}
\rho_0(\vec{r}) = \phi_i(\vec{r})\phi_i(\vec{r})
\end{equation}
and have the property that the resulting correlation energy~(\ref{Ec})
is equal to the exact one. It is obvious that the correction potentials thus defined
are not unique. The simplest form would be a constant shift
(independent of $\vec{r}$) for either
all occupied or all virtual energy levels: $u_{\rm o}(\vec{r})=u[\rho_0]$,
$u_{\rm v}=0$ or vice versa, but in that case the density functional
$u[\rho_0]$ should be highly non-local and very hard to model without
violating the size-consistency of an approximate method. One would be interested
in the models where these potentials were local in nature.

For the full-CI wavefunction of the form
\begin{equation}
\label{tCI}
\Psi_{\rm CI} =
 \Phi + T_i^a \Phi_i^a + \frct14 T_{ij}^{ab} \Phi_{ij}^{ab}
 + \frct1{36} T_{ijk}^{abc} \Phi_{ijk}^{abc} + \dots
\end{equation}
the correlation energy can be expressed as~(\ref{Eijab}) in terms of
the doubles coefficients $T_{ij}^{ab}$ alone, so it would be natural to require
the equations~(\ref{tijab}) to generate the values $\tau_{ij}^{ab}$
which approximate $T_{ij}^{ab}$ in~(\ref{tCI}).
Having only $n_{\rm o}(n_{\rm o}+1)/2 + n_{\rm v}(n_{\rm v}+1)/2 - 1$
matrix elements $u_i^j$ and $u_a^b$ as parameters
($n_{\rm o}$ and $n_{\rm v}$ are the dimensions of the occupied and virtual
subspaces, the latter can be in principle infinite)
it is not possible, in general, to reproduce all
$n_{\rm o}(n_{\rm o} - 1)n_{\rm v}(n_{\rm v} - 1)/4$ nontrivial
values of $T_{ij}^{ab}$. The differences between the reduced one-electron
quantities
\begin{eqnarray}
Q_i^j &=& \tau_{ik}^{ab} \tau_{jk}^{ab} - T_{ik}^{ab} T_{jk}^{ab}, \\
Q_a^b &=& \tau_{ij}^{ac} \tau_{ij}^{bc} - T_{ij}^{ac} T_{ij}^{bc}
\end{eqnarray}
can be brought to zero by an appropriate choice of $u_i^j$ and $u_a^b$
in~(\ref{tijab}), in that case, however, the correlation energy~(\ref{Ec})
may differ from the exact one. The minimization of either
\begin{equation}
q = Q_i^j Q_i^j + Q_a^b Q_a^b
\end{equation}
or 
\begin{equation}
q = Q_i^j \bar{R}_{ik}^{jl} Q_k^l + Q_a^b \bar{R}_{ac}^{bd} Q_c^d
\end{equation}
with respect to $u_i^j$ and $u_a^b$
with the constraint of reproducing the full-CI correlation energy by~(\ref{Eijab})
leads to the equations which uniquely define the correction potentials
in either case. A more simple function
\begin{equation}
q =
\left(\tau_{ij}^{ab} - T_{ij}^{ab}\right)
\left(  (F_c^a\delta_{bd} + F_d^b\delta_{ac})\delta_{ik}\delta_{jl}
      - \delta_{ac}\delta_{bd}(F_i^k\delta_{jl} + F_j^l\delta_{ik})
\right)
\left(\tau_{kl}^{cd} - T_{kl}^{cd}\right)
\end{equation}
can also be used as a measure of the deviation between the doubles coefficients.

The existence of such (not necessarily multiplicative) potentials
$u_{\rm o}(\vec{r};[\rho_0])$ and
$u_{\rm v}(\vec{r};[\rho_0])$
 as functionals of the Hartree-Fock density
follows from the fact that there is a one-to-one mapping between
the external potential~$v(\vec{r})$ and the Hartree-Fock density~$\rho_0(\vec{r})$,
the relationship can be schematically described as
$\rho_0 \Rightarrow v \Rightarrow \Psi_{\rm CI}
 \Rightarrow u_{\rm o}, u_{\rm v}$.

In analogy to the traditional density-functional theory
the development of approximate functionals of this new type can
start from the consideration of the uniform electron gas
as the simplest model system. A local density approximation of the form
\begin{eqnarray}
\label{LDAo}
u_{\rm o}(\vec{r}) &=&
u^{\rm ueg}\left(\vp\rho_0(\vec{r})\right)
\alpha\left(\vp\rho_0(\vec{r})\right), \\
\label{LDAv}
u_{\rm v}(\vec{r}) &=&
u^{\rm ueg}\left(\vp\rho_0(\vec{r})\right)
\left(1+\alpha\left(\vp\rho_0(\vec{r})\right)\right),
\end{eqnarray}
can be proposed,
where the function $u^{\rm ueg}(\rho)$ is chosen as to reproduce,
 within this method,
the exact correlation energy of the uniform electron gas,
and the function $\alpha(\rho)$ can be parametrized in some way, the simplest
choices being the constants $\alpha=0$ or $\alpha=-1$.

\section{Conclusions}

The new form of the correlation energy functional of the Hartree-Fock density
proposed here is based on a physically meaningful description
of the electron correlation using the two-particle equations~(\ref{tijab}).
Approximate methods of this type should be able to describe
the dispersion interaction in a natural way
and should not contain self-interaction error.
It remains to be seen if the simplest local density
approximation~(\ref{LDAo}) and~(\ref{LDAv}) will show higher accuracy
in molecular applications than the standard MP2 method.
Such comparison will become possible as soon as the function~$u^{\rm ueg}(\rho)$
will be available from numerical calculations of the uniform electron gas,
which we are currently investigating.
More sophisticated density-functional models of
our corrections potentials
$u_{\rm o}(\vec{r};[\rho_0])$ and
$u_{\rm v}(\vec{r};[\rho_0])$
can also be developed, an empirical parametrization~\cite{B97}
being the simplest choice.

\end{document}